\documentclass[12pt]{article}
\usepackage{epsfig,amssymb,amsmath,psfrag,subfigure,rotate,color}

\allowdisplaybreaks
\newcommand{\insertfig}[2]{\includegraphics[width=#1cm]{#2}}

\def \be  {\begin{equation}}
\def \ee  {\end{equation}}
\def \ba  {\begin{eqnarray}}
\def \ea  {\end{eqnarray}}
\def \baa {\begin{eqnarray*}}
\def \eaa {\end{eqnarray*}}
\def \lab #1 {\label{#1}}

\newcommand\re[1]{(\ref{#1})}
\def\d{\hbox{{d}\kern-.20em\hbox{l}}}

\def \matrix #1 {\left(\begin{array}{cc} #1 \end{array}\right)}

\def \res{\mathop{\rm res}\nolimits}

\newcommand \vev [1] {\langle{#1}\rangle}

\newcommand{\ft}[2]{{\textstyle\frac{#1}{#2}}}
\begin{document}

\begin{titlepage}

\thispagestyle{empty}

\vspace*{3cm}

\centerline{\large \bf Descent Equation for superloop and cyclicity of OPE}
\vspace*{1cm}

\centerline{\sc A.V.~Belitsky}

\vspace{10mm}

\centerline{\it Department of Physics, Arizona State University}
\centerline{\it Tempe, AZ 85287-1504, USA}

\vspace{2cm}

\centerline{\bf Abstract}

\vspace{5mm}

We study the so-called Descent, or $\bar{Q}$, Equation for the null polygonal supersymmetric Wilson loop in the framework of the pentagon operator
product expansion. To properly address this problem, one requires to restore the cyclicity of the loop broken by the choice of OPE channels. In the 
course of the study, we unravel a phenomenon of twist enhancement when passing to a cyclically shifted channel. Currently, we focus on the consistency 
of the all-order Descent Equation for the particular case relating the NMHV heptagon to MHV hexagon. We find that the equation establishes a relation 
between contributions of different twists and successfully verify it in perturbation theory making use of available bootstrap predictions for elementary pentagons.

\end{titlepage}

\setcounter{footnote} 0

\newpage

\pagestyle{plain}
\setcounter{page} 1


\newpage

\section{Introduction}

The superamplitude $\mathcal{A}_N$ in planar $\mathcal{N} = 4$ superYang-Mills theory is known to be dual to the superWilson loop
\cite{Alday:2007hr,Drummond:2007cf,Brandhuber:2007yx,CaronHuot:2010ek,Mason:2010yk,Belitsky:2011zm}
\begin{align}
\mathcal{W}_N = \exp\left( i g \oint_{C_N} \mathbb{A} \right)
\, ,
\end{align}
defined by a  superconnection $\mathbb{A}$ residing on a piecewise light-like contour $C_N$ in chiral superspace. The $\mathcal{W}_N$, being an off-shell correlator, 
provides a fully nonperturbative description of $\mathcal{A}_N$. What makes this correspondence even more powerful is that $\mathcal{W}_N$ can be systematically 
analyzed in the multi-collinear regimes, i.e., when certain adjacent links become parallel \cite{Alday:2010ku,Gaiotto:2011dt,Belitsky:2011nn,Sever:2012qp}. This 
expansion receives a rigorous nonperturbative reincarnation within the so-called pentagon operator product expansion (OPE) \cite{Basso:2013vsa}. The power of 
the latter lies in the fact that all of its ingredients can be computed to all orders in 't Hooft coupling making use of the hidden integrablility of the theory 
\cite{Basso:2013vsa,Basso:2010in,Basso:2011rc,Fioravanti:2013eia,Basso:2013pxa,Basso:2013aha,Basso:2014koa,Belitsky:2014lta,Belitsky:2015efa,Basso:2014hfa}. 
In spite of the fact that there is a plethora of data on scattering amplitudes that heavily relies on ordinary and dual superconformal symmetries 
\cite{Drummond:2008vq,Mason:2009qx,Arkani-Hamed:2013jha,CaronHuot:2011kk}, the above formalism obscures the most basic symmetries such as supersymmetry, 
cyclicity etc. 

The chiral nature of the superWilson loop representation itself, while preserves some tree-level dual symmetries, masks others and makes them coupling 
dependent in spite of the fact that there are no intrinsic short-distance anomalies associated with them\footnote{Of course, there are generators that do develop
true anomalies due to ultraviolet divergencies, like conformal boost etc.}. One particular generator that received a close
attention in this regard was the Poincar\'e supersymmetry $\bar{Q}$ \cite{CaronHuot:2011kk,Bullimore:2011kg}, i.e., the chiral conjugate of $Q$. Its action on the 
superloop was cast in an all-loop conjecture\footnote{To date, the Descent Equation viewed as a Ward identity of the antichiral supersymmery generator has evaded 
rigorous studies due to lack of a proper regularization scheme that leaves manifest symmetries of the superloop intact. The main culprit in these analyses is the correlation 
functions of field equations of motion with superholonomies and strong UV divergencies associated with light-cone nature of the contour \cite{Belitsky:2011zm}.} 
\cite{CaronHuot:2011kk} that was dubbed the Descent Equation, see Eq.\ \re{QbarEq} below. Its power was uncovered in the fact that it mixes different orders in 
perturbative series, enabling one to predict higher loop amplitudes from the ones an order lower. Of particular importance for this application was the cyclicity of the 
loop that provided contributions adding up together to yield correct final answer. As a consequence, the goal of the current study will be twofold. We will unravel how the cyclic 
contributions are implemented in the Descent Equation from the point of view of OPE. And then we will realize what the Descent Equation imply for the mixing of different 
twists in the operator series.

Our subsequent presentation is organized as follows. In the following section, we recap the form of the equation for the finite Wilson loop observables which
are natural from the point of view of the pentagon OPE. Next, we provide a preliminary discussion of the equation relating one-loop NMHV heptagon to 
two-loop MHV hexagon at leading twist in flux-tube excitations. As we observe there, to properly incorporate cyclic contributions we have, in principle, to
resum the entire OPE series. To start with, we go beyond leading twist and uncover the form of the Wilson loop at twist two in Sections \ref{SectionGpsi} and \ref{SectionHpsi}. 
To guide ourselves in the quest of uncovering cyclicity, we use available exact one-loop expressions for the heptagon and conjecture the form of cyclic contributions
in terms of OPE data. Making use of all-order predictions, we then verify that it is indeed correct by going to one loop order higher when all genuine two-particle states start
to contribute. Finally we conclude.

\section{Collinear limit and Descent Equation}

A natural observable from the point of view of OPE is a properly subtracted superWilson loop $\mathcal{W}_N$. It is related to to the
ratio function $\mathcal{R}_N = \mathcal{A}_N/A_N^{\rm BDS}$ that was devised in Ref.\ \cite{CaronHuot:2011kk} according to
\begin{align}
\mathcal{W}_N = \mathcal{R}_N W_N^{\rm U(1)}
\, .
\end{align}
Here $\ln W_N^{\rm U(1)}$ is the sum of connected correlators of Wilson loops in U(1) theory between reference squares in a chosen tessellation of the $N$-gon with 
the coupling constant $g^2_{\scriptscriptstyle\rm U(1)}$ replaced by (one quarter of) the exact cusp anomalous dimensions in $\mathcal{N} = 4$ theory, 
$\Gamma (g) = 4 g^2 - 8 g^4 \zeta_2 + \dots$,
\begin{align}
\ln W_N^{\rm U(1)} = \frac{1}{4} \Gamma (g) X
\, .
\end{align}
The function $X$ depends on $3N-5$ conformal cross ratios $X = X (u_1, \dots , u_{3N-5})$. The superloop admits a terminating expansion in Grassmann variables,
\begin{align}
\mathcal{W}_N = \sum_{n=0}^{N-4} \mathcal{W}_{N,n}
\, ,
\end{align}
with each term being a degree-$4n$ polynomials. They correspond to N$^{n}$MHV amplitudes, up to an overall factor of the 't Hooft coupling, namely, $\mathcal{W}_{N, n} 
= g^{2 n} \mathcal{A}_{N, n}$.

The action of the $\bar{Q}$-operator,
\begin{align}
\bar{Q}^A_a = \sum_{n=1}^N \chi_n^A \frac{\partial}{\partial Z_n^a}
\, ,
\end{align}
on the $N$-point N$^{n}$MHV observable can be cast in the form
\begin{align}
\label{QbarEq}
\bar{Q}^A_\alpha  \mathcal{W}_{N,n} 
= 
\frac{\Gamma (g)}{4 g^2}
\sum_i^{N + 1} \int d^{2|3} \mathcal{Z}^A_{\alpha \, i} 
\left[
\mathcal{W}_{N+1,n+1}
-
\mathcal{W}^{\rm tree}_{N+1,1}
\mathcal{W}_{N,n}
\right]
\, ,
\end{align}
where in the right-hand side one takes a collinear limit of an $N+1$ point N$^{n+1}$MHV Wilson loop. Notice an extra factor of $1/g^2$ in the above equation that arises
due to the aforementioned conversion from amplitudes to Wilson loops. The limit is accomplished by means of a proper parametrization of the near-collinear expansion of adjacent sites
parametrized by supertwistors $\mathcal{Z}_i = (Z_i, \chi_i)$ built from momentum twistors $Z_i$ \cite{Hodges} and their Grassmann counterparts $\chi_i$. A particularly convenient 
form is gained in the OPE framework by encoding all inequivalent polygons with the action of symmetries of intermediate squares, see Appendix \ref{AppendixPolygons}. 
For the case at hand, the supersymmetric collinear limit emerges from the relation
\begin{align}
\label{CollinearLimitZ}
\mathcal{Z}^{(n)}_1 
= 
\mathcal{Z}^{(n)}_n 
- 
{\rm e}^{- \tau^\prime} \mathcal{Z}^{(n)}_{n-1} 
+ 
{\rm e}^{-\tau^\prime + 2 \sigma^\prime} (1 + {\rm e}^{- \tau^\prime - \sigma^\prime + i \phi^\prime}) \mathcal{Z}^{(n)}_2 
+ 
{\rm e}^{- 2 \tau^\prime + \sigma^\prime + i \phi^\prime} \mathcal{Z}^{(n)}_3
\, ,
\end{align}
and subsequently taking $\tau^\prime \to \infty$. This implies an expansion at the bottom of the polygon in terms of flux-tube excitations of increasing twists. The measure 
$d^{2|3} \mathcal{Z}_1$ \cite{CaronHuot:2011kk}, however,
\begin{align}
d^{2|3} \mathcal{Z}_{\alpha \, 1}^A =   \oint_{|\varepsilon| = 0_+} \frac{d \varepsilon^\prime \varepsilon^\prime}{2 \pi i} \int d {\rm e}^{2 \sigma^\prime}
\int (d^3 \chi_1)^A \, \bar{n}_\alpha
\, ,
\end{align}
--- where we defined $\varepsilon^\prime = {\rm e}^{- \tau^\prime}$ and $ \bar{n}_\alpha = \varepsilon_{\alpha\beta\gamma\delta} Z^\beta_{n-1} Z^\gamma_n Z^\delta_1$,
--- singles out one flux-tube fermion.

\section{Preliminaries on Descent Equation}

The right-hand side of the Descent Equation projects out a single fermionic excitation on the bottom of the polygon, while the top can
absorb any multi-particle states with fermionic quantum numbers. Let us, however, start our analysis by considering its left-hand side.
We will focus on the MHV hexagon as a case of study. At leading twist, it receives a contribution from a single gauge field created from the vacuum in the
operator channel chosen by the parametrization of the momentum twistors introduced in the Appendix \ref{AppendixPolygons},
\begin{align}
\mathcal{W}_{6,0} = 1 + ({\rm e}^{i \phi} + {\rm e}^{- i \phi}) {\rm e}^{- \tau} \mathcal{W}_{6 [1](2)} + \dots
\, ,
\end{align}
with\footnote{\label{FootnoteLabels} Here and below, we accept a notation that the subscripts in $\mathcal{W}_{N (t_1, \dots, t_{3N-5}) [h_1, \dots, h_{3N-5}]}$ stand for the
numbers of cusps in the superloop $N$, with twists $t_1, \dots$ of excitations propagating on sequential intermediate squares and their corresponding total
double helicities $h_1, \dots$.} \cite{Basso:2013aha}
\begin{align}
\label{HexagonGluon}
\mathcal{W}_{6 [1](2)} = \int_{\mathbb{R}} d \mu_{\rm g} (v)
\, .
\end{align}
Here we used a compound notation for the differential measure of the ${\rm p}$-type flux-tube excitation along with the propagating phase encoded by
its energy $E_{\rm p}$ and momentum $p_{\rm p}$,
\begin{align}
d \mu_{\rm p} (v)
= 
\frac{d v}{2 \pi} \mu_{\rm p} (v) {\rm e}^{- \tau [E_{\rm p} (v) - 1] + i \sigma p_{\rm p} (v)}
\end{align}

For brevity, we select $\alpha = 4$ component of the $\bar{Q}^A_\alpha$ generator. Its action can be easily evaluated on the twist-one hexagon to read
\begin{align}
\label{QW6}
\bar{Q}^A_4 {\rm e}^{- \tau} \mathcal{W}_{6[1](2)}
=
\chi_4^A
{\rm e}^{- \tau}
\int_{\mathbb{R}} d \mu_{\rm g} (v) 
\ft12
\big(
E_{\rm g} (v) + i p_{\rm g} (v)
\big)
+
\dots
\, ,
\end{align}
where ellipses stand for cyclic contributions accompanied by other Grassmann variables. Expanding the measure,
$\mu_{\rm p} = g^2 \mu^{(1)}_{\rm p} + g^4 \mu^{(2)}_{\rm p} + \dots$, the energy and momentum, $E_{\rm p} = 1 + g^2 E^{(1)}_{\rm p} + \dots$ and
$p_{\rm p} = 2 u + g^2 p^{(1)}_{\rm p} + \dots$ in perturbative series, we can shift the integration contour in the lower half of the complex rapidity plane, 
$v \to v - \ft{i}{2}$ and rewrite the result in the form
\begin{align}
\label{QbarW6}
\bar{Q}^A_4 {\rm e}^{- \tau} \mathcal{W}_{6[1](2)}
=
&
-
\chi_4^A
{\rm e}^{- \tau}
\int_{\mathbb{R} + i 0} \frac{dv}{2 \pi} {\rm e}^{2 i v \sigma}
\bigg\{
g^2  {\rm e}^\sigma \mu^{(1)}_{\rm F} (v) 
\\
&
+ 
g^4
\left[
{\rm e}^\sigma
\left(
\mu^{(2)}_{\rm F} (v) + ( i \sigma p^{(1)}_{\rm F} (v) - \tau E_{\rm F}^{(1)} (v) ) \mu^{(1)}_{\rm F} (v) 
\right)
-
\left( 2 \tau + 2 \sigma - i  p^{(1)}_{\rm g} (v) \right) \mu^{(1)}_{\rm g} {(v)}
\right]
\nonumber\\
&
\qquad\qquad\qquad\qquad\qquad\qquad\qquad\qquad\qquad\qquad\quad \ \, 
+
O (g^6)
\bigg\}
+
\dots
\, , \nonumber
\end{align}
at the lowest two orders of perturbation theory. To arrive at this expression we used the following 
results. First, it is immediate to demonstrate that\footnote{Cf.\ these to the relations $E_{\rm F} (v) - E_{\rm h} (v - \ft{i}{2}) = i p_{\rm f} (v)$ and $p_{\rm F} (v) - p_{\rm h} (v - 
\ft{i}{2}) = i E_{\rm f} (v)$ found in Ref.\ \cite{Basso:2011rc}.}
\begin{align}
\label{EPgEPFfrelations}
E_{\rm g} (v - \ft{i}{2}) = E_{\rm F} (v) - i p_{\rm f} (v - i)
\, , \qquad
p_{\rm g} (v - \ft{i}{2}) = p_{\rm F} (v) - i E_{\rm f} (v - i)
\, ,
\end{align}
by confirming these identities order-by-order in 't Hooft coupling. For the measures, we have in the lowest few orders
\begin{align}
(i - v) \mu_{\rm g}^{(1)} (v - \ft{i}{2}) 
&
= i \mu_{\rm F}^{(1)} (v)
\, , \nonumber\\
(i - v) \mu_{\rm g}^{(2)} (v - \ft{i}{2}) 
&
= i \mu_{\rm F}^{(2)} (v) - \ft{i}{2} E^{(1)}_{\rm g} (v - \ft{i}{2}) \mu_{\rm g}^{(1)} (v - \ft{i}{2}) + \frac{ \pi (1 + 2 i v)}{v^2 (v - i)^2 \sinh (\pi v)}
\, \dots \nonumber
\end{align}

\begin{figure}[t]
\begin{center}
\mbox{
\begin{picture}(0,200)(100,0)
\put(0,-150){\insertfig{17}{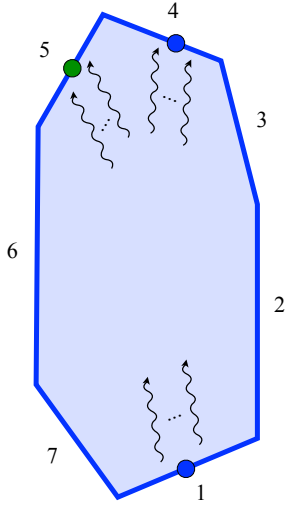}}
\end{picture}
}
\end{center}
\caption{ \label{OPEchannelsHeptagon} Two OPE channels relevant for the Descent Equation involving heptagon. The two channels
are obtained from each other by a mirror reflection with respect to the line going through $Z_2$ and connecting to the vertex 
$Z_4 \wedge Z_5$.}
\end{figure}

A naked eye inspection of  Eq.\ \re{QbarW6} then immediately suggests that the first line and the first term in the second line emerge from a single-fermion 
exchange, as anticipated from the Descent Equation. This can be easily verified by computing the $\sigma^\prime$ integral of the $\chi_1^3 \chi_4$ 
component \re{W741111} of the heptagon, see Fig.\ \ref{OPEchannelsHeptagon}, which is extracted by the $d^{2|3} \mathcal{Z}_1$ measure. It yields to the
lowest two orders in $g^2$
\begin{align}
\int d \sigma^\prime \, \mathcal{W}^{(4)}_{7 [1,1] (1,1)} (\sigma^\prime, \sigma)
=
&
- {\rm e}^\sigma
\int \frac{dv}{2 \pi} {\rm e}^{2 i u v}
\bigg\{
g^2 \mu_{\rm F}^{(1)} (v)
\\
+
g^4
\Big[
\mu^{(2)}_{\rm F} (v) 
&
+
( i \sigma p^{(1)}_{\rm F} (u) - \tau E_{\rm F}^{(1)} (u)
+
2 \zeta_2
) 
\mu^{(1)}_{\rm F} (u)
\Big]
+ O (g^6)
\bigg\}
\, , \nonumber
\end{align}
where $\sigma$ is associated with the rapidity $v$ and $\sigma^\prime$ with $u$, respectively. The $\zeta_2$ term gets cancelled upon multiplication by the
cusp anomalous dimension thus providing agreement alluded to above. 

However, the second term in the square brackets of Eq.\ \re{QbarW6} is much more enigmatic. Obviously, there is yet another term that is projected out by the
fermionic measure from the heptagon, namely, at leading twist one finds an extra $\chi_5$ term in addition to already addressed $\chi_4$,
\begin{align}
\int d^{2|3} \mathcal{Z}_1 \mathcal{W}_{7,1}
&
=
{\rm e}^{ - \tau} 
\int d \sigma^\prime
\left(
\chi_4^A \mathcal{W}^{(4)}_{7 [1,1] (1,1)}
+
\chi_5^A \mathcal{W}^{(5)}_{7 [1,1] (1,1)}
\right)
+ \, {\rm higher \ twists}
\, , \nonumber
\end{align} 
So the $\chi_5$ term should be the source of the remaining last term in Eq.\ \re{QbarW6}, which does not admit an obvious OPE interpretation. Thus, one has 
to understand how it changes as we apply the cyclic permutation to it, i.e., the one that turns $\chi_5 \to \chi_4$. It appears though that one needs to restore the exact 
dependence on the cross ratios before moving to a different channel. Is it only possible provided one resums the entire OPE series in this channel? As we will 
demonstrate below, it is not quite the case but one has to move beyond leading twist in the adjacent channel to induce a leading contribution after the cyclic shift.

\section{Fermionic heptagon}

As we stated above, we have to unravel the structure of subleading terms in the OPE expansion of fermionic components of the heptagon. Thus 
we turn to a thorough analysis of $\chi_1^3 \chi_4$ and $\chi_1^3 \chi_5$ Grassmann structures with emphasis on the lowest twist contribution at the
bottom and twist-two on the top. The general form of this expansion takes the form
\begin{align}
\label{GenericW7}
\mathcal{W}_{7,1} 
= 
\sum_{j = 4,5}
\chi_1^3 \chi_j \sum_{n_1, n_2} \sum_{h_1, h_2} {\rm e}^{- n^\prime \tau^\prime - n \tau} {\rm e}^{ i (h^\prime \phi^\prime + h \phi)/2} 
\mathcal{W}^{(j)}_{[n^\prime, n] (h^\prime, h)} (\sigma^\prime, \sigma; g)
+
\dots
\, ,
\end{align}
where the dots stand for irrelevant Grassmann structures and the nomenclature for the labels was explained in the footnote \ref{FootnoteLabels}.

\subsection{Twist-one: fermion exchange}

\begin{figure}[t]
\begin{center}
\mbox{
\begin{picture}(0,280)(200,0)
\put(0,-200){\insertfig{25}{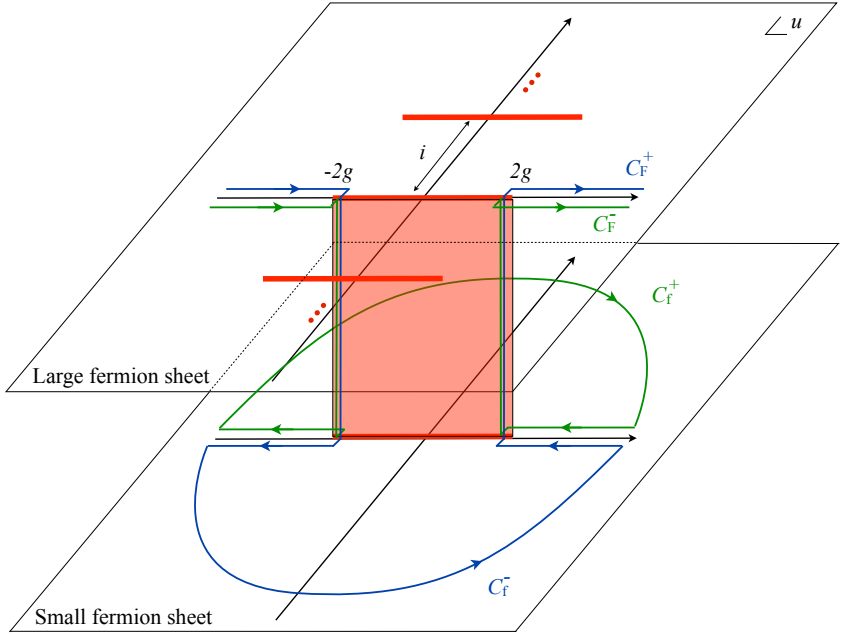}}
\end{picture}
}
\end{center}
\caption{ \label{FermionicContours} The large and small fermion complex planes are glued together along the square root branch cut on the real axis
$[-2g, 2g]$ (shown by the bold red interval) into a two-sheeted Riemann surface. The integration contours for the fermions in the OPE expressions are 
shown for the $\chi_4$ in blue, $\mathbb{C}^+_\Psi = \mathbb{C}^+_{\rm F} + \mathbb{C}^-_{\rm f}$, and for $\chi_5$ one in green, $\mathbb{C}^-_\Psi 
= \mathbb{C}^-_{\rm F} + \mathbb{C}^+_{\rm f}$, paths.}
\end{figure}

To start with, we will recall the leading effect from the twist-one fermion propagating in the OPE channels in question \cite{Belitsky:2014lta}, i.e.,
proportional to  ${\rm e}^{- \tau^\prime - \tau} {\rm e}^{i \phi^\prime/2 + i \phi/2}$. Both $\chi_1^3 \chi_4$ and $\chi_1^3 \chi_5$ Grassmann structures are 
cumulatively given by
\begin{align}
\label{OPEOneFermion}
\mathcal{W}^{(j)}_{7 [1,1](1,1)} (\sigma^\prime, \sigma)
=
\int_{\mathbb{C}^+_\Psi} d \mu_\Psi (u_1) x[u_1] \int_{\mathbb{C}^{(j)}_\Psi} d \mu_\Psi (v_1) P_{\Psi | \Psi} (- u_1 | v_1)
\, . 
\end{align}
They are expressed in terms of the helicity non-flip fermionic pentagon transition $P_{\Psi | \Psi}$ \cite{Basso:2014koa} with the measure of the initial-state 
fermion accompanied by a helicity form factor given by the Zhukowski variable $x[u] = \ft12 ( u + \sqrt{u^2 - (2 g)^2} )$. Above, the integration 
contours are shown in Fig.\ \ref{FermionicContours}. The one for the fermion on the bottom of the heptagon is $\mathbb{C}^+_\Psi 
= \mathbb{C}^+_{\rm F} + \mathbb{C}^-_{\rm f}$ with $\mathbb{C}^+_{\rm F} = \mathbb{R} + i 0$ and $\mathbb{C}^-_{\rm f}$ running a half-circle 
in the lower semiplane of the complex plane. The top contour depends on the supertwistor which the flux-tube excitation is ``attached'' to. For the $\chi_1^3 \chi_4$
channel it is the same $\mathbb{C}^{(4)}_\Psi = \mathbb{C}^+_\Psi$, while for $\chi_1^3 \chi_5$, it is ``flipped'' on the Riemann surface with respect to the imaginary
axis, i.e., $\mathbb{C}^{(5)}_\Psi = \mathbb{C}^-_\Psi = \mathbb{C}^-_{\rm F} + \mathbb{C}^+_{\rm f}$. It can be explained by the fact that the latter channel is a
mirror reflection of the original $\chi_1^3 \chi_4$ one, thus in a given tessellation it corresponds to a different collinear limit.

In perturbation theory, since there are no poles in the integrand, only the large fermion contributes to the OPE, yielding
\begin{align}
\label{W741111}
\mathcal{W}^{(4,5)}_{7 [1,1](1,1)} (\sigma^\prime, \sigma)
=
\int_{\mathbb{R} + i 0} d \mu_{\rm F} (u) x[u]  \int_{\mathbb{R} \pm i 0} d \mu_{\rm F} (v) P_{\rm F|F} (- u| v)
\, .
\end{align}
Its expansion in 't Hooft coupling can be easily verified to agree with explicit amplitudes by using, for instance, the package of Ref.\ \cite{Bourjaily:2013mma}

\subsubsection{Integral of single-particle contribution}
\label{FermionIntegralSection}

The Descent Equation involves an integral over the position $\sigma^\prime$. A perturbative analysis to a very high order reveals that the result is given by
\begin{align}
\label{W74twist1}
\Gamma (g) \int d \sigma^\prime \, \mathcal{W}^{(4)}_{7 [1,1](1,1)} (\sigma^\prime, \sigma)
=
- 2 i g^2 
 \int_{\mathbb{R} + i 0} d \mu_\Psi (v_1)
\, ,
\end{align}
where the prefactor of the exact cusp anomalous dimension arises from the integral
\begin{align}
\int d u_1 \, \mu_\Psi (u_1) {\rm e}^{- \tau^\prime [E_\Psi (u_1) - 1]}
x[u_1] \delta \big(p_\Psi (u_1)\big) 
P_{\Psi | \Psi} (- u_1 | v_1) 
=
- \frac{2 i g^2}{\Gamma (g)}
\, .
\end{align}
Superficially the left-hand side depends on the rapidity $v_1$, but in reality it is only a function of the 't Hooft coupling. Notice that the dependence on the cross-ratio 
$\tau^\prime$ trivializes in light of the fact that the fermion mass is one at any value of the coupling \cite{Alday:2007mf},
\begin{align} 
p_\Psi (u) = 0
\, , \qquad
E_\Psi (u) = 1
\, .
\end{align}

For the $\chi_1^3 \chi_5$ channel, the fermion contour in the final state runs below the real axis. However, due to the Fourier exponent it has to be closed in the 
upper half plane. As a result of moving the contour just above the real axis, one picks up a pole at $v_1 = 0$ on the real axis. The latter term induces a divergent
contribution when integrated with respect to $\sigma^\prime$. Namely, we find that it is coupling independent
\begin{align}
\res\limits_{v_1 = 0} P_{\rm F|F} (-u_1|v_1) \mu_{\rm F} (v_1) {\rm e}^{ - \tau [ E_{\rm F} (v_1) - 1]+ i \sigma p_{\rm F} (v_1)}
=
- i 
\, ,
\end{align}
such that the two contributions differ by a single-fermion exchange in the NMHV hexagon
\begin{align}
\label{W4minus41111}
\mathcal{W}^{(5)}_{7 [1,1](1,1)}  (\sigma^\prime, \sigma) = \mathcal{W}^{(4)}_{7 [1,1](1,1)}  (\sigma^\prime, \sigma) + \mathcal{W}_{6 [1](1)} (\sigma^\prime)
\, ,
\end{align}
with
\begin{align}
\mathcal{W}_{6 [1](1)} 
=
- i
\int_{\mathbb{R} + i 0} d \mu_{\rm F} (u_1) x[u_1] 
\, .
\end{align}
It is the last term in the right-hand side that diverges when integrated with respect to $\sigma^\prime$. This contribution gets subtracted in the Descent Equation 
\re{QbarEq} by the last term in its right-hand side.

\subsection{Twist-two: fermion-gluon in final state}
\label{SectionGpsi}

We now turn to twist-two effects. As exhibited by Eq.\ \re{GenericW7}, at twist-two there is a contribution that enters with the helicity prefactor ${\rm e}^{3i\phi/2}$. It 
corresponds to a fermion-gluon pair absorbed by the top portion of the Wilson loop in a given OPE channel. Its all-order expression in coupling constant reads\footnote{
All pentagon transitions used here and below can be found in Refs.\ \cite{Belitsky:2014sla,Belitsky:2014lta,Belitsky:2015efa}.}
\begin{align}
\mathcal{W}^{(j)}_{7 [1,2](1,3)} 
=
g \int_{\mathbb{C}^+_\Psi} d \mu_\Psi (u_1) 
\int_{\mathbb{C}^{(j)}_\Psi} d \mu_\Psi (v_1) \int_{\mathbb{R}} d \mu_{\rm g} (v_2)
\frac{
x[u_1] 
P_{\Psi | {\rm g}} (- u_1| v_2) P_{\Psi | \Psi} (- u_1| v_1)
}{
\sqrt{x^+[v_2] x^- [v_2]} P_{ {\rm g} | \Psi} (v_2 | v_1) P_{ {\rm g} | \Psi} (- v_2 | - v_1)
} 
\, ,
\end{align}
where we used the factorized form of one-to-two particle transition pentagon \cite{Belitsky:2014lta} and fermion-gluon absorption form factor \cite{Belitsky:2015efa}.
The helicity form factor is expressed in terms of shifted Zhukowski variables $x^\pm [u] \equiv x [u^\pm]$ where $u^\pm = u \pm \ft{i}{2}$. The bottom fermion resides 
on the large sheet, while the one on the top can be split in the above formula into the small (f) and large (F) contributions
\begin{align}
\mathcal{W}^{(j)}_{7 [1,2](1,3)} = \mathcal{W}^{(j)}_{7 \, {\rm F|fg}} + \mathcal{W}^{(j)}_{7 \, {\rm F|Fg}} 
\, .
\end{align}

We start with $j=4$ case first. At lowest two orders in coupling, only the small fermion contributes to the Wilson loop and induces a nontrivial effect that reads
\begin{align}
\mathcal{W}^{(4)}_{7 \, {\rm F|fg}} 
=
g \int_{\mathbb{R} + i 0} d \mu_{\rm F} (u_1) x[u_1] \int_{\mathbb{R} + i 0} d \mu_{{\rm gf}} (v_2) 
\left[ \frac{x^- [v_2]}{x^+ [v_2]} \right]^{1/2} P_{{\rm F} | {\rm g}} (- u_1| v_2) P_{{\rm F} | {\rm f}} (- u_1| v_2^-) 
\, ,
\end{align}
where we introduced the small-fermion--gluon measure \cite{Belitsky:2014lta},
\begin{align}
\mu_{\rm gf} (v_2) = 
\res\limits_{v_1 = v_2^-}
\frac{g^2 \mu_{\rm f} (v_1) \mu_{\rm g} (v_2)}{x [v_1] P_{\rm f|g} (v_1|v_2) P_{\rm f|g} (- v_1|- v_2)}
\, .
\end{align}
At $O(g^6)$ and higher, $\mathcal{W}_{7 [1,2](1,3)}$ receives an additional term from the large fermion
\begin{align}
\mathcal{W}^{(4)}_{7 \, {\rm F|Fg}} 
=
g \int_{\mathbb{R} + i 0} d \mu_{\rm F} (u_1) 
\int_{\mathbb{R} + i 0} d \mu_{\rm F} (v_1) \int_{\mathbb{R} + i 0} d \mu_{\rm g} (v_2)
\frac{
x[u_1] 
P_{{\rm F} | {\rm g}} (- u_1| v_2) P_{{\rm F} | {\rm F}} (- u_1| v_1)
}{
\sqrt{x^+[v_2] x^- [v_2]} P_{ {\rm g} | {\rm F}} (v_2 | v_1) P_{ {\rm g} | {\rm F}} (- v_2 | - v_1)
} 
\, .
\end{align}

Similar analysis can be performed for $j = 5$. The differences in the fermion contour result in differences of various contributions. The small 
fermion now reads instead
\begin{align}
\mathcal{W}^{(5)}_{7 \, {\rm F|fg}} 
=
g \int_{\mathbb{R} + i 0} d \mu_{\rm F} (u_1) x[u_1] \int_{\mathbb{R}} d \, \widetilde\mu_{{\rm gf}} (v_2) 
\left[ \frac{x^+ [v_2]}{x^- [v_2]} \right]^{1/2} P_{{\rm F} | {\rm g}} (- u_1| v_2) P_{{\rm F} | {\rm f}} (- u_1| v_2^+) 
\, ,
\end{align}
where compared to the previous equation, since the pole $v_2 = v_1^+$ was picked up in the upper half plane of the lower Riemann sheet, the 
composite measure has changed,
\begin{align}
\widetilde\mu_{\rm gf} (v_2) 
= 
\res\limits_{v_1 = v_2^+}
\frac{g^2 \mu_{\rm f} (v_1) \mu_{\rm g} (v_2)}{x [v_1] P_{\rm f|g} (v_1|v_2) P_{\rm f|g} (- v_1|- v_2)}
\, ,
\end{align}
as well as the square root prefactor was flipped. As above, at order $g^6$ and beyond, the Wilson loop gets a new term from the large fermion
\begin{align}
\mathcal{W}^{(5)}_{7 \, {\rm F|Fg}} 
=
g \int_{\mathbb{R} + i 0} d \mu_{\rm F} (u_1) 
\int_{\mathbb{R} - i 0} d \mu_{\rm F} (v_1) \int_{\mathbb{R} + i 0} d \mu_{\rm g} (v_2)
\frac{
x[u_1] 
P_{{\rm F} | {\rm g}} (- u_1| v_2) P_{{\rm F} | {\rm F}} (- u_1| v_1)
}{
\sqrt{x^+[v_2] x^- [v_2]} P_{ {\rm g} | {\rm F}} (v_2 | v_1) P_{ {\rm g} | {\rm F}} (- v_2 | - v_1)
} 
\, ,
\end{align}
which differs from $\chi_4$ by the position of the contour in the large fermion rapidity complex plane. The correctness of these expression can be verified by
comparing them in the perturbative expansion with the results of Ref.\ \cite{Bourjaily:2013mma}. It would be important to extend it to higher orders, especially in a 
fully analytic manner relying on the methods of Ref.\ \cite{Papathanasiou:2013uoa}, using the heptagon bootstrap program 
\cite {Golden:2014xqa,Golden:2014xqf,Golden:2014pua,Drummond:2014ffa} that generalizes earlier results on the hexagon \cite{Dixon:2011nj,Dixon:2014iba} .

\subsubsection{Integral of gluon-fermion contributions}

The calculation of the $\sigma^\prime$ integrals of contributions introduced in the previous section follows the same route as for single fermion in Section 
\ref{FermionIntegralSection}, but now we have to evaluate 
the rapidity integral involving two pentagons. For the $j=4$ contribution, this can be successfully accomplished order by order in perturbation theory and then 
resummed back into an exact function of the 't Hooft coupling constant. For the integral involving the small fermion, we find accordingly
\begin{align}
\int d u_1 \, \mu_{\rm F} (u_1) {\rm e}^{- \tau^\prime [E_{\rm F} (u_1) - 1]}
x[u_1] \delta \big(p_{\rm F} (u_1)\big) 
P_{\rm F | f} (- u_1 | v_2^-) P_{\rm F | g} (- u_1 | v_2)
=
- \frac{2 i g^3}{\Gamma (g)} \frac{1}{\sqrt{x^+[v_2] x^- [v_2]}} 
\, .
\end{align}
For the one with the large fermion, the expression for the right-hand side looks identical
\begin{align}
\int d u_1 \, \mu_{\rm F} (u_1) {\rm e}^{- \tau^\prime [ E_{\rm F} (u_1) - 1]}
x[u_1] \delta \big(p_{\rm F} (u_1)\big) 
P_{\rm F | F} (- u_1 | v_1) P_{\rm F | g} (- u_1 | v_2)
=
- \frac{2 i g^3}{\Gamma (g)} \frac{1}{\sqrt{x^+[v_2] x^- [v_2]}} 
\, .
\end{align}
However, a simple counting of the powers of 't Hooft coupling for the large fermion demonstrates that its contribution it
postponed one order higher than the integrand itself since the leading term in its expansion vanishes after the $u_1$ integration. Thus, 
the large fermion appears accompanied by a gluon starting only from three loops in the Descent Equation.

For $j = 5$, in complete analogy with the single fermion, the $\sigma^\prime$ integral turns out to be divergent. So it requires a subtraction. To make it as transparent 
as possible, let us calculate the difference between the $\chi_5$ and $\chi_4$ contributions first. A careful all-order perturbative analysis demonstrates that 
the latter can be expressed in terms of the following expression
\begin{align}
\label{DiffW541213}
\mathcal{W}^{(5)}_{7 [1,2](1,3)} - \mathcal{W}^{(4)}_{7 [1,2](1,3)} 
= 
\Delta \mathcal{W}_{7 [1,2](1,3)} 
\end{align}
with
\begin{align}
\Delta \mathcal{W}_{7 [1,2](1,3)} 
\equiv
\frac{1}{g}
\int_{\mathbb{R} + i 0}  d \mu_{\rm F} (u_1)
\int_{\mathbb{R} + i 0} d \mu_{\rm g} (v_2) \, x [u_1] P_{{\rm F}|{\rm g}} (- u_1 | v_2) \sqrt{x^-[v_2] x^+ [v_2]}
\, . 
\end{align}
It is important to emphasize that, starting from three-loop order, only the total sum of small and large fermion contributions becomes $u_1$
independent as shown by the factor accompanying $x\, \mu_{\rm F} P_{{\rm F}|{\rm g}}$ in the integrand of the above equation!

To proceed further, we form the NMHV ratio functions,
\begin{align}
\mathcal{P}^{(j)}_{7[1,2](1,3)} (\sigma^\prime, \sigma)
=
\mathcal{W}^{(j)}_{7[1,2](1,3)} (\sigma^\prime, \sigma)
-
\mathcal{W}^{(j)}_{7[1,1](1,1)} (\sigma^\prime, \sigma) 
\mathcal{W}_{6[1](2)} (\sigma)
\, , 
\end{align}
with the gluon flux-tube excitation propagating on the bosonic hexagon $\mathcal{W}_{6[1](2)}$, see Eq.\ \re{HexagonGluon}.
Substituting Eq.\ \re{W4minus41111} into above \re{DiffW541213}, we find
\begin{align}
\mathcal{P}^{(5)}_{7[1,2](1,3)} (\sigma^\prime, \sigma)
=
\mathcal{P}^{(4)}_{7[1,2](1,3)} (\sigma^\prime, \sigma)
+
\left[
\Delta \mathcal{W}_{7 [1,2](1,3)} 
-
\mathcal{W}_{6[1](1)} (\sigma_1) \mathcal{W}_{6[1](2)} (\sigma)
\right]
\, .
\end{align}
The integral of the regularized expression is finite and can be cast in a concise form,
\begin{align}
\int d \sigma^\prime \big[
\Delta \mathcal{W}_{7 [1,2](1,3)} (\sigma^\prime, \sigma)
&
-
\mathcal{W}_{6[1](1)} (\sigma^\prime) \mathcal{W}_{6[1](2)} (\sigma)
\big]
\\
&
= \frac{i g^2}{\Gamma (g)}
\int_{\mathbb{R} + i 0} d \mu_{\rm g} (v_2) \left[ x^- [v_2] - \frac{g^2}{x^+[v_2]} - \ft{i}{2} \left( E_{\rm g} (v_2) + i p_{\rm g} (v_2) \right) \right]
\, . \nonumber
\end{align}
This concludes our discussion of integrals involving fermion-gluon pairs in the OPE of the heptagon.

\subsection{Twist-two: antifermion-scalar in final state}
\label{SectionHpsi}

Finally, we address the $\mathcal{W}^{(j)}_{7 [1,2](1,-1)}$. A simple counting of quantum numbers immediately suggests that there are two additive contributions,
one from the hole-antifermion and another from antigluon-fermion pair,
\begin{align}
\mathcal{W}^{(j)}_{7 [1,2](1,-1)}
=
\mathcal{W}^{(j)}_{7 \, \Psi |\bar{\Psi} {\rm h}} +  \mathcal{W}^{(j)}_{7 \, \Psi | \Psi \bar{\rm g}} 
\, .
\end{align}
These admit a representation in terms of pentagons as follows,
\begin{align}
\mathcal{W}^{(j)}_{7 \, \Psi |\bar{\Psi} {\rm h}}
&
=
\frac{2}{g^3}
\int_{\mathbb{C}_\Psi} d \mu_\Psi (u_1)
\\
&
\times
\int_{\mathbb{C}_\Psi^{(j)}} d \mu_\Psi (v_1)
\int_{\mathbb{R}} d \mu_{\rm h} (v_2)
\frac{x^{3/2} [u_1]  x [v_1]  P_{\Psi|\bar\Psi} (- u_1 | v_1) P_{\Psi | {\rm h}} (- u_1 | v_2) 
}{
(v_1 - v_2 - \ft{i}{2})(v_1 - v_2 + \ft{3 i}{2})P_{\Psi | {\rm h}} (v_1 | v_2) P_{ \Psi | {\rm h}} (- v_1 | - v_2)}
\, , \nonumber\\
\label{W7121m1Generic}
\mathcal{W}^{(j)}_{7 \, \rm F|F \bar{g}} 
&
=
g
\int_{\mathbb{C}_\Psi} d \mu_{\rm F} (u_1) 
\\
&
\times
\int_{\mathbb{C}_\Psi^{(j)}} d \mu_{\rm F} (v_1)
\int_{\mathbb{R} + i 0} d \mu_{\rm g} (v_2) 
\frac{}{} 
\frac{
x[u_1] \sqrt{x^+[v_2] x^- [v_2]} P_{\Psi|\Psi} (- u_1| v_1) P_{\Psi|\bar{\rm g}} (- u_1| v_2)
}{
(v_1 - v_2 - \ft{i}{2}) x [v_1]  P_{\Psi | \bar{\rm g}} (v_1| v_2) P_{\Psi | \bar{\rm g}} (- v_1| - v_2)}
\, . \nonumber
\end{align}

Having these generic expressions, we can rewrite them in specific OPE channels accounting for the difference in the fermionic contours. For $j=4$ channel, 
decomposing the fermion into small and large contributions, we obtain
\begin{align}
\label{chi4fermiholeLower}
\mathcal{W}^{(4)}_{7 [1,2](1,-1)}
=
\mathcal{W}^{(4)}_{7 \, \rm F|\bar{f} h} + \mathcal{W}^{(4)}_{7 \, \rm F|\bar{F} h} + \mathcal{W}^{(5)}_{7 \, \rm F|F \bar{g}} 
\, ,
\end{align}
The first term in the right-hand side starts at order $g^2$ and induces a tree-level term in the amplitude. The large antifermion--hole sets in an order
higher, i.e., $O (g^4)$, while $\mathcal{W}^{(4)}_{7 [1,2](1,-1)}$ starts receiving contributions from large-fermion--antigluon pair from two loops.
They read individually, 
\begin{align}
\mathcal{W}^{(4)}_{7 \, \rm F|\bar{f} h} 
&
=
g 
\int_{\mathbb{R} + i 0} d \mu_{\rm F} (u_1) 
\int_{\mathbb{R}}  d \mu_{\rm h} (v_2) \mu_{\rm f} (v_2 - \ft{3 i}{2})
\frac{x^{3/2} [u_1] P_{\rm F|\bar{f}} (- u_1 | v_2 - \ft{3 i}{2}) P_{\rm F|h} (- u_1 | v_2) }{ x[v_2 - \ft{3 i}{2}] P_{\rm f|h} (v_2 - \ft{3 i}{2} | v_2) P_{\rm f|h} (- v_2 + \ft{3 i}{2} | - v_2)}
\, , \\
\label{W74FFh}
\mathcal{W}^{(4)}_{7 \, \rm F|\bar{F} h} 
&
=
\frac{2}{g^3}
\int_{\mathbb{R} + i 0} d \mu_{\rm F} (u_1) 
\nonumber\\
&
\times
\int_{\mathbb{R} + i 0} d \mu_{\rm F} (v_1)
\int_{\mathbb{R}} d \mu_{\rm h} (v_2) 
\frac{
x^{3/2}[u_1] x [v_1]  P_{\rm F|\bar{F}} (- u_1 | v_1) P_{\rm F| \rm h}  (-u_1 | v_2)
}{
(v_1 - v_2 - \ft{i}{2}) (v_1 - v_2 + \ft{3 i}{2}) P_{\rm F|h} (v_1 | v_2) P_{\rm F|h} (- v_1 | - v_2)}
\, ,
\\
\label{W74FFgbar}
\mathcal{W}^{(4)}_{7 \, \rm F|F \bar{g}} 
&
=
g
\int_{\mathbb{R} + i 0} 
d \mu_{\rm F} (u_1) 
\\
&
\times
\int_{\mathbb{R} + i 0}
d \mu_{\rm F} (v_1)
\int_{\mathbb{R} + i 0}
d \mu_{\rm g} (v_2) 
\frac{
x[u_1] \sqrt{x^+[v_2] x^- [v_2]} P_{\rm F|F} (- u_1| v_1) P_{\rm F|\bar{g}} (- u_1| v_2)
}{
(v_1 - v_2 - \ft{i}{2}) x [v_1] P_{\rm F|\bar{g}} (v_1| v_2) P_{\rm F|\bar{g}} (- v_1| - v_2)}
\, . \nonumber
\end{align}

For $\chi_1^3 \chi_5$ channel, all one has to do is to use the corresponding fermion contour. Then one immediately realizes that compared to the
previously studied sector, there will be an additional contribution from the small-fermion--antigluon in Eq.\ \re{W7121m1Generic} due to the location 
of the simple pole above the real axis where we close our integration contour. Then we have
\begin{align}
\mathcal{W}^{(5)}_{7 [1,2](1,-1)}
=
\mathcal{W}^{(5)}_{7 \, \rm F|\bar{f} h} + \mathcal{W}^{(5)}_{7 \, \rm F|\bar{F} h} + \mathcal{W}^{(5)}_{7 \, \rm F|f \bar{g}} + \mathcal{W}^{(5)}_{7 \, \rm F|F \bar{g}} 
\, .
\end{align}
To the lowest two orders of perturbation theory only small fermions contribute to the right-hand side as we now explain. Namely, closing the integration contour in the upper
half plane in $\mathcal{W}_{\Psi |\bar{\Psi} {\rm h}}$ yields the result $\mathcal{W}^{(5)}_{\rm F|\bar{f} h}$
\begin{align}
\label{chi4fermiholeUpper}
\mathcal{W}^{(5)}_{7 \rm F|\bar{f} h}
=
g
\int_{\mathbb{R} + i 0} d \mu_{\rm F} (u_1) 
\int_{\mathbb{R}}  d \mu_{\rm h} (v_2) \mu_{\rm f} (v_2^+)
\frac{x^{3/2} [u_1] P_{\rm F|\bar{f}} (- u_1 | v_2^+) P_{\rm F|h} (- u_1 | v_2) }{ x^+ [v_2] P_{\rm f|h} (v_2^+ | v_2) P_{\rm f|h} (- v_2^+ | - v_2)}
\, .
\end{align}
The large-fermion is determined by the integrand of Eq.\ \re{W74FFh} with the $v_2$ integral running just below the real axis, $\mathbb{R} - i 0$. This contribution
vanishes at $O (g^4)$. At this order the small fermion in the pair with antigluon
\begin{align}
\label{Wchi5PsiPsibarG}
\mathcal{W}^{(5)}_{7 \, \rm F|f \bar{g}} 
=
\frac{1}{g}
\int_{\mathbb{R} + i 0} 
d \mu_{\rm F} (u_1) x[u_1]
\int_{\mathbb{R}}
d \mu_{\rm g} (v_2) \mu_{\rm f} (v_2^+) x^+[v_2]
\sqrt{x^+[v_2] x^- [v_2]} \frac{P_{\rm F|f} (- u_1| v_2^+) P_{\rm F|\bar{g}} (- u_1| v_2)}{P_{\rm f|\bar{g}} (v_2^+| v_2) P_{\rm f|\bar{g}} (- v_2^+| - v_2)}
\end{align}
starts at one loop, postponing the effect  of the large fermion to two loops. The latter is determined by the same equation as in \re{W74FFgbar}, where one
has to shift the integration contour with respect to $v_2$ in the lower half-plane. 

Expansion in 't Hooft coupling allows us to confirm these predictions at lowest two orders with explicit amplitudes \cite{Bourjaily:2013mma}.

\subsubsection{Integral of antifermion-hole and fermion-antigluon contributions}

To uncover contributions of the above twist-two effects to the Descent Equation, we have to finally evaluate the $\sigma^\prime$ integrals. Again we start with the 
convergent $j = 4$ operator channel. To this end, we need the following set of rapidity integrals involving the small antifermion and the hole,
\begin{align}
\label{Chi4hfbarIntegral}
\int d u_1 \, \mu_{\rm F} (u_1) {\rm e}^{- \tau^\prime [ E_{\rm F} (u_1) - 1]}
x[u_1] \delta \big(p_{\rm F} (u_1)\big) 
P_{\rm F | \bar{f}} (- u_1 | v_2 - \ft{3i}{2} ) P_{\rm F | h} (- u_1 | v_2)
=
\frac{2 g^3}{\Gamma (g)} 
\, ,
\end{align}
the large antifermion and the hole
\begin{align}
\label{Chi4hFbarIntegral}
\int d u_1 \, \mu_{\rm F} (u_1) {\rm e}^{- \tau^\prime [ E_{\rm F} (u_1) - 1]}
x[u_1] \delta \big(p_{\rm F} (u_1)\big) 
P_{\rm F | \bar{F}} (- u_1 | v_1) P_{\rm F | h} (- u_1 | v_2)
=
\frac{2 g^3}{\Gamma (g)} 
\, ,
\end{align}
and last but not least, the large fermion and antigluon
\begin{align}
\int d u_1 \, \mu_{\rm F} (u_1) {\rm e}^{- \tau^\prime [ E_{\rm F} (u_1) - 1]}
x[u_1] \delta \big(p_{\rm F} (u_1)\big) 
P_{\rm F | F} (- u_1 | v_1) P_{\rm F | \bar{g}} (- u_1 | v_2)
=
- \frac{2 g}{\Gamma (g)} \sqrt{x^+[v_2] x^- [v_2]}
\, .
\end{align}

Now we move to the $j=5$ case. First for antifermion-hole contribution $\mathcal{W}^{(5)}_{7 \, \Psi|\bar\Psi {\rm h}}$, we find that the integral involving the small antifermion 
is identical to Eq.\ \re{Chi4hfbarIntegral}. In fact, these two are particular cases of a more general formula for a generic value of $v_1$ in $P_{\rm F | \bar{f}} (- u_1 | v_1 )$. 
Next, for the large fermion we can use Eq.\ \re{Chi4hFbarIntegral} in spite of the fact that the integration contour for the outgoing fermion lies below the real axis. The reason 
for this is that while crossing the real axis one acquires a pole along the way, this term is not singular for $u_1 = 0$. Finally, we turn to the antigluon-fermion final state, 
$\mathcal{W}^{(5)}_{7 \, \Psi|\Psi \bar{\rm g}}$. In this case, the $\sigma^\prime$ integral is not convergent, so one has to form the ratio function and thus subtract a factorized 
contribution,
\begin{align}
\mathcal{P}^{(5)}_{7[1,2](1,-1)} (\sigma^\prime, \sigma)
=
\mathcal{W}^{(5)}_{7[1,2](1,-1)} (\sigma^\prime, \sigma)
-
\mathcal{W}^{(5)}_{7[1,1](1,1)} (\sigma^\prime, \sigma) \mathcal{W}_{6[1](2)} (\sigma)
\, . 
\end{align}
Substituting \re{W4minus41111}, we can split the result into two terms,
\begin{align}
\mathcal{W}^{(5)}_{7 \, \Psi|\Psi \bar{\rm g}} (\sigma^\prime, \sigma)
&
- 
\mathcal{W}^{(5)}_{7[1,1](1,1)} (\sigma^\prime, \sigma) \mathcal{W}_{6[1](2)} (\sigma)
\\
&
=
\mathcal{W}^{(4)}_{7 \, \Psi|\Psi \bar{\rm g}} (\sigma^\prime, \sigma)
-
\mathcal{W}^{(4)}_{7[1,1](1,1)} (\sigma^\prime, \sigma) \mathcal{W}_{6[1](2)} (\sigma)
\nonumber\\
&\qquad\qquad\qquad\quad \ \ \,
+
\left[
\Delta \mathcal{W}^{(5)}_{7 \, \Psi|\Psi \bar{\rm g}} (\sigma^\prime, \sigma)
-
\mathcal{W}_{6[1](1)} (\sigma^\prime) \mathcal{W}_{6[1](2)} (\sigma)
\right]
\, , \nonumber
\end{align}
with the first two terms in its right-hand side addressed in previous sections. The integral of the square bracket can be expressed in a concise form
\begin{align}
\int_0^\infty d \sigma^\prime
\bigg[
\Delta \mathcal{W}^{(5)}_{7 \, \Psi|\Psi \bar{\rm g}} (\sigma^\prime, \sigma)
&
-
\mathcal{W}_{6[1](1)} (\sigma^\prime) \mathcal{W}_{6[1](2)} (\sigma)
\bigg]
\\
&
=
\frac{i g^2}{\Gamma (g)}
\int d \mu_{\rm g} (v_2)
\left[
- x^- [v_2] + \frac{g^2}{x^+ [v_2]} - \ft{i}{2} \left( E_{\rm g} (v_2) + i p_{\rm g} (v_2) \right)  
\right]
\, . \nonumber
\end{align}

One can combine all of the above ingredients together and substitute them into the Descent Equation. We will not give the cumulative formula here to 
save space. It is obvious from the representation of these results, that the overall power of the inverse cusp anomalous dimension cancels against the one
in the right-hand side of Eq.\ \re{QbarEq}. Thus, this proves the all-order form of the Descent Equation provided that we can establish an agreement between
OPE series on both of its sides. This is what we will accomplish next getting some inspiration from perturbation theory.

\subsection{Cyclic permutation at one loop}

Having analyzed all twist-two flux-tube excitations in the two channels of interest, we have to find a way to cyclically shift $j = 5$ channel down to $j=4$.
Since we currently lack a dynamical understanding of this mechanism from the point of view of underlying flux-tube dynamics, we will choose
a more pragmatic route. As a guidance, we will start with an exact expression for the one-loop heptagon and work out the Descent Equation for it. 
According to Eq.\ \re{QbarEq}, the action of the $\bar{Q}$ generator on the two-loop bosonic hexagon yields \cite{CaronHuot:2011kk}
\begin{align}
\bar{Q}^A_\alpha \mathcal{W}_{6,0}
&
= 
 \bar{Q}^A_\alpha \ln \frac{\vev{6724}}{\vev{6723}}
\int_0^\infty d t \, 
I^{(4)} (t | u_1, u_2, u_3, v)
\\
&
+
 \bar{Q}^A_\alpha \ln \frac{\vev{6725}}{\vev{6723}}
\int_0^\infty d t \, 
I^{(5)} (t | u_1, u_2, u_3, v)
+
\, 
{\rm cyclic}
\, , \nonumber
\end{align}
where the right-hand side receives contributions from the $\chi_1^3 \chi_4$ and $\chi_1^3 \chi_5$ structures, in the first and second term, respectively, 
and corresponds in the OPE language to a single fermion at the bottom of the heptagon and all twists absorbed at the top. The collinear expansion on the 
top of the heptagon admits a systematic classification within the pentagon framework, namely, we immediately find
\begin{align}
{\rm e}^{2 \sigma^\prime} I_4 ({\rm e}^{2 \sigma^\prime} | u_1, u_2, u_3, v)
=
&
-
{\rm e}^{- \tau}
\mathcal{P}^{(4)}_{7 \, [1,1](1,1)} (\sigma^\prime, \sigma, \tau^\prime = 0, \tau) 
\\
&
-
{\rm e}^{- 2 \tau}
\left(
{\rm e}^{i \phi} \mathcal{P}^{(4)}_{7 \, [1,2](1,3)} + {\rm e}^{- i \phi} \mathcal{P}^{(4)}_{7 \, [1,2](1,-1)} 
\right)
(\sigma^\prime, \sigma, \tau^\prime = 0, \tau)
+
O ({\rm e}^{- 3 \tau})
\, , \nonumber\\
{\rm e}^{2 \sigma^\prime} I_5 ({\rm e}^{2 \sigma^\prime} | u_1, u_2, u_3, v)
=
&
-
{\rm e}^{- \tau}
\mathcal{P}^{(5)}_{7 \, [1,1](1,1)} (\sigma^\prime, \sigma, \tau^\prime = 0, \tau) 
\\
&
-
{\rm e}^{- 2 \tau}
\left(
{\rm e}^{i \phi} \mathcal{P}^{(5)}_{7 \, [1,2](1,3)} + {\rm e}^{- i \phi} \mathcal{P}^{(5)}_{7 \, [1,2](1,-1)} 
\right)
(\sigma^\prime, \sigma, \tau^\prime = 0, \tau)
+
O ({\rm e}^{- 3 \tau})
\, , \nonumber
\end{align}
where the remaining dependence of the ratio functions on the $\tau$ and $\tau^\prime$ is polynomial due to higher order corrections to eigen-energies of
flux-tube excitations. Above, we set $\tau^\prime$ to zero everywhere since these contributions vanish after $\sigma^\prime$ integration owing to the 
Goldstone theorem \cite{Alday:2007mf}.

Finally, we have to move the $\chi_5$ contribution to the $\chi_4$ channel. This is achieved by a cyclic shift of twistors that results in a
change of cross ratios as shown in Eqs.\ \re{ShftedUs}. We will use the exact expression for $I_5$ \cite{CaronHuot:2011kk},
\begin{align}
I_5 (t | u_1, u_2, u_3, v)
&=
-
\frac{u_3}{t (u_3 + t)}
\bigg[
-
\ln (1 + t) \ln \frac{1 + t}{t}
-
\ln \frac{u_3 (1 + t)}{u_3 + t} \ln \frac{u_2 (u_3 + t)}{t}
\\
&
+
{\rm Li}_2 (1 - u_2) + {\rm Li}_2 (1 - u_3)
-
{\rm Li}_2 \left( \frac{1 - u_3}{1 + t} \right) + {\rm Li}_2 \left(1 - \frac{u_2}{1 + t} \right) 
\bigg]
\nonumber\\
&+
\frac{1}{t (u_3 + t)}
\bigg[
\ln \frac{u_3 (u_3 + t)}{u_3 + t} \ln \frac{u_3 + t}{t}
+ 
{\rm Li}_2 (1 - u_1)
-
{\rm Li}_2 \left( 1- \frac{u_1 t_1}{u_3 + t} \right)
\bigg]
\nonumber\\
&
+ \frac{v - u_3}{1 + t}
\bigg[
\ln \frac{u_2}{1+t} \ln \frac{u_1 t}{u_3 + t}
+
{\rm Li}_2 \left( 1 - \frac{u_2}{1 + t}  \right)
+
{\rm Li}_2 \left( 1 - \frac{u_1 t}{u_3 + t} \right)
-
\zeta_2
\bigg]
\, . \nonumber
\end{align}
Then the transition to the cyclic channel, is achieved by replacing the conformal cross ratios $u \to \tilde{u}$ summarized in Appendix \ref{AppendixPolygons}. Expanding 
the result at $\tau \to \infty$, we deduce the following expression after integration with respect to $\sigma^\prime$
\begin{align}
\int_{-\infty}^\infty d \sigma^\prime \, {\rm e}^{2 \sigma^\prime} I_5 ({\rm e}^{2 \sigma^\prime} | \tilde{u}_1, \tilde{u}_2, \tilde{u}_3, \tilde{v})
=
0 
&
-
{\rm e}^{- \tau}
\int_{-\infty}^\infty d \sigma^\prime
\bigg[
i 
\left(
\mathcal{P}^{(5)}_{7 \, [1,2](1,3)} - \mathcal{P}^{(5)}_{7 \, [1,2](1,-1)}
\right) \sin \phi
\\
+
\Big(
\mathcal{P}^{(4)}_{7 \, [1,2](1,-1)} + \mathcal{P}^{(4)}_{7 \, [1,2](1,3)}
&-
\mathcal{P}^{(5)}_{7 \, [1,2](1,3)} - \mathcal{P}^{(5)}_{7 \, [1,2](1,-1)}
+
{\rm e}^{\sigma} \mathcal{P}^{(4)}_{7 \, [1,1](1,1)}
\Big)
\cos \phi
\bigg]
+
O({\rm e}^{- 2 \tau})
\, . \nonumber
\end{align}
Here the leading-twist effect vanishes upon integration! The subleading terms are expressed in terms of OPE exchanges worked out above for the $\chi_4$ and $\chi_5$ 
Grassmann components. The parity-even term here defines the leading contribution to the $\chi_4$-channel of the Descent Equation after the cyclic shift. This is the effect 
of twist enhancement we alluded to in the Introduction. This expression can be cast in a very concise form at this order in the coupling, namely,
\begin{align}
\label{Twis2ToTwist1}
2 \int_{-\infty}^\infty d \sigma^\prime
\Big(
\mathcal{P}^{(4)}_{7 \, [1,2](1,-1)} + \mathcal{P}^{(4)}_{7 \, [1,2](1,3)}
&-
\mathcal{P}^{(5)}_{7 \, [1,2](1,3)} - \mathcal{P}^{(5)}_{7 \, [1,2](1,-1)}
+
{\rm e}^{\sigma} \mathcal{P}^{(4)}_{7 \, [1,1](1,1)}
\Big)
\\
&
=
-
g^4
\int_{\mathbb{R} + i 0} \frac{d v_1}{2 \pi} {\rm e}^{2 i v_1 \sigma}
\mu_{\rm g}^{(1)} (v_1)
\left(
2 \tau + 2 \sigma -  i p_{\rm g}^{(1)} (v_1)
\right)
+
O(g^6)
\, . \nonumber
\end{align}
As we can see, the $O(g^4)$ contribution is in agreement with the last term in Eq.\ \re{QbarW6}. It is important to realize, however, that at one-loop the left-hand 
side is not unique. Namely, there is a  relation between $\mathcal{W}^{(4)}_{7 [1,2](1,-1)}$  and $\mathcal{W}^{(4)}_{7 [1,2](1,3)}$ components of the superloop,
\begin{align}
\label{chi4121m1Relation}
\mathcal{W}^{(4)}_{7 [1,2](1,-1)}
=
\mathcal{W}^{(4)}_{7 [1,2](1,3)}
+
{\rm e}^{\sigma}
\mathcal{W}^{(4)}_{7 [1,1](1,1)}
+
O (g^6)
\, , 
\end{align} 
which is valid to order $g^4$ only and thus implies that one can replace $\mathcal{P}^{(4)}_{7 \, [1,2](1,-1)} + \mathcal{P}^{(4)}_{7 \, [1,2](1,3)}
+ {\rm e}^{\sigma} \mathcal{P}^{(4)}_{7 \, [1,1](1,1)} \to 2 \mathcal{P}^{(4)}_{7 \, [1,2](1,-1)}$ in the above equation. Taken at its face value,
this relation yields an incorrect OPE representation for the $\mathcal{P}^{(4)}_{7 \, [1,2](1,-1)}$ since in the lowest two orders of perturbation theory 
the left-hand side can be cast in the form \re{chi4fermiholeUpper}. Obviously, the pole at $v_2 = v_1^+$ is not at the right side of the
real axis to be naturally accommodated into OPE. There is an immediate problem that one needs to reconcile, namely, how the two results, Eq.\ 
\re{chi4fermiholeUpper} and \re{chi4fermiholeLower} can be compatible. The first one has only small fermion contribution while the latter one has 
both, large and small. It turns out that if one ignores the proper choice of the contour and uses $\mathbb{C}^-_\Psi$ instead of $\mathbb{C}^+_\Psi$, then 
$\mathbb{C}^+_{\rm f}$ is closed in the upper half plane and one picks up a pole at  $v_2 = v_1^+$ and gets the result given in Eq.\ 
\re{chi4fermiholeUpper}. At the same time, as we already mentioned before, the genuine two-particle twist-two contribution (that sets in at order $g^4$) 
vanishes, so one is left with small-fermion--hole pair alone as found in the relation \re{chi4121m1Relation}.

Therefore, to put the result \re{Twis2ToTwist1} on a firmer foundation, we compared the left- and right-hand sides of the Descent Equation at $O(g^6)$, 
when all two-particle excitations contribute to the OPE. As a result, we confirmed the agreement, i.e., the equality of Eq.\ \re{QW6} to the sum of Eq.\ \re{W74twist1} 
and \re{Twis2ToTwist1} multiplied by the factor of the cusp anomalous dimension. This is the main result of this paper.

\section{Conclusions}

In this work, we analyzed the Descent Equation within the OPE. We demonstrated that the factor of the cusp anomalous dimensions naturally arises 
in its right-hand side from the pentagon framework confirming in this manner the all loop structure of the Descent Equation. We have established a 
phenomenon of twist enhancement as one passes from direct to a cyclic channel. Namely, the leading  effect emerges from the subleading contribution in the 
adjacent channel. Added up with the twist one excitations in the direct channel, it proves the consistency of the OPE expansion with the $\bar{Q}$-equation. 
It would be interesting to extend this consideration to event higher twists and other nonMHV polygons.

It is very important to understand the twist enhancement from the underlying flux-tube picture. A handwaving argument along the lines, that one can move 
the fermionic excitation on the top of the polygon from one site to another making use of a ``mirror" transformation for the fermion which increases its twist 
and promotes it to a string of two excitation separated in the rapidity space \cite{Basso:2014koa}, is too vague and imprecise. Moreover, from the point of
view of explicit calculations, this could only account for some of the contributions.

As a next step it is interesting to focus on the structure of the equation at strong  coupling\footnote{See recent work on resumption of the OPE series and its agreement 
with TBA \cite{Fioravanti:2015dma}}. Contrary to weak coupling, as $g \to \infty$ the large fermion decouples since its energy and momentum scale with $g$, while the 
small fermion?s mass remains unrenormalized. A quick look at the left-hand side of the Descent Equation demonstrates that the structure of propagating exponent is very suggestive 
if one shifts the integration contour and uses the all-order relations \re{EPgEPFfrelations}. However to correctly reproduce the prefactor, one has to understand how the 
cyclic channel gets added to the total result. A rather subtle issue to be addressed is about composite states involving a fermion an arbitrary number of massless (at 
strong coupling) holes that contribute on equal footing with a single small fermion, see, e.g., Ref.\ \cite{Basso:2014jfa}. These questions will be addressed in a future 
publication.

\section*{Acknowledgments}

We would like to thank Jacob Bourjaily for very instructive correspondence and the organizers of the ``Flux-tube Workshop" for hospitality at Perimeter Institute 
at the final stages of this work. This research was supported by the U.S. National Science Foundation under the grants PHY-1068286 and PHY-1403891.

\appendix

\section{Parametrization of polygons}
\label{AppendixPolygons}

Making use of the projective invariance of momentum twistors, we will use the following rescaled version of the later
for the hexagon
\begin{align}
Z_1^{(6)} 
&
= \left({\rm e}^{- \tau + 2 \sigma}, 0, {\rm e}^{\sigma + i \phi}, {\rm e}^{- 2 \tau + \sigma + i \phi} \right)
\, , \\
Z_2^{(6)} 
&
= \left(1,0,0,0\right)
\, , \nonumber\\
Z_3^{(6)} 
&
= \left(-1,0,0,1\right)
\, , \nonumber\\
Z_4^{(6)} 
&
= \left(0, 1, -1, 1 \right)
\, , \nonumber\\
Z_5^{(6)} 
&
= \left(0,1, 0, 0\right)
\, , \nonumber\\
Z_6^{(6)} 
&
= \left(0, {\rm e}^{- \tau}, {\rm e}^{\sigma + i \phi}, 0 \right)
\, , \nonumber
\end{align}
and heptagon
\begin{align}
Z_1^{(7)} 
&
= \left({\rm e}^{- \tau^\prime + 2 \sigma^\prime}, 0, {\rm e}^{\sigma^\prime + i \phi^\prime}, {\rm e}^{- 2 \tau^\prime + \sigma^\prime + i \phi^\prime} \right)
\, , \\
Z_2^{(7)} 
&
= \left(1,0,0,0\right)
\, , \nonumber\\
Z_3^{(7)} 
&
= \left(-1,0,0,1\right)
\, , \nonumber\\
Z_4^{(7)} 
&
= \left(- {\rm e}^{- \tau}, {\rm e}^{- \sigma - i \phi}, - {\rm e}^{- \sigma - i \phi}, {\rm e}^{- \sigma - i \phi} (1 +  {\rm e}^{- 2 \tau}) + {\rm e}^{- \tau} \right)
\, , \nonumber\\
Z_5^{(7)} 
&
= \left(0, {\rm e}^{ - \sigma - i \phi} + {\rm e}^{- \tau - 2 \sigma}, - {\rm e}^{- \sigma - i \phi},  {\rm e}^{ - \sigma - i \phi}  \right)
\, , \nonumber\\
Z_6^{(7)} 
&
= \left(0,1,0,0 \right)
\, , \nonumber\\
Z_7^{(7)} 
&
= \left(0, {\rm e}^{- \tau^\prime}, {\rm e}^{\sigma^\prime + i \phi^\prime}, 0 \right)
\, , \nonumber
\end{align}
respectively, compared to Refs.\ \cite{Basso:2013aha,Belitsky:2015efa}. These are very well suited for the collinear expansion within the framework of the Descent Equation, 
in particular ensuring Eq.\ \re{CollinearLimitZ}.

The collinear limit $Z_1 \to Z_7$ at the bottom of  the heptagon leaves just three conformal cross ratios analogous to the one of the hexagon
and one non-spacetime cross ratio
\begin{align}
u_1 
&
=
\left. \frac{(2,3,4,5)(5,6,7,2)}{(2,3,5,6)(4,5,7,2)} \right|_{\tau^\prime \to \infty}
=
\frac{{\rm e}^{i \phi}}{{\rm e}^{2 \sigma + i \phi} + {\rm e}^{- \tau + \sigma} + {\rm e}^{- \tau + \sigma + 2 i \phi} + {\rm e}^{i \phi} (1 + {\rm e}^{-2 \tau})}
\, , \nonumber\\
u_2 
&
= 
\left. \frac{(3,4,5,6)(6,7,2,3)}{(3,4,6,7)(2,3,5,6)}  \right|_{\tau^\prime \to \infty}
=
\frac{{\rm e}^{- 2 \tau}}{1 + {\rm e}^{- 2 \tau}}
\, , \nonumber\\
u_3 
&
= 
\left. \frac{(4,5,6,7)(7,2,3,4)}{(4,5,7,2)(3,4,6,7)}  \right|_{\tau^\prime \to \infty}
=
\frac{{\rm e}^{2 \sigma}}{1 + {\rm e}^{- 2 \tau}} u_1
\, ,  \nonumber\\
v 
&
= 
\left. \frac{(6,7,3,5)(7,2,3,4)}{(6,7,3,4)(7,2,3,5)}  \right|_{\tau^\prime \to \infty}
=
\frac{{\rm e}^{\sigma}}{(1 + {\rm e}^{- 2 \tau}) ({\rm e}^{\sigma} + {\rm e}^{- \tau + i \phi})}
\, . \nonumber
\end{align}
We also introduce cyclically shifted cross ratios $i \to i+1$,
\begin{align}
\label{ShftedUs}
\tilde{u}_1 
&
=
\left. \frac{(3,4,5,6)(6,7,2,3)}{(3,4,6,7)(5,6,2,3)} \right|_{\tau^\prime \to \infty}
=
u_2
\, , \\
\tilde{u}_2 
&
= 
\left. \frac{(4,5,6,7)(7,2,3,4)}{(4,5,7,2)(3,4,6,7)}  \right|_{\tau^\prime \to \infty}
=
u_3
\, , \nonumber\\
\tilde{u}_3 
&
= 
\left. 
\frac{
(5,6,7,2)(2,3,4,5)
}{
(5,6,2,3)(4,5,7,2)
}  
\right|_{\tau^\prime \to \infty}
=
u_1
\, ,  \nonumber\\
\tilde{v} 
&
= 
\left. \frac{(7,2,4,6)(2,3,4,5)}{(7,2,4,5)(2,3,4,6)}  \right|_{\tau^\prime \to \infty}
=
\left(
1 + {\rm e}^{- \tau + \sigma + i \phi} + {\rm e}^{- 2 \tau}
\right)
u_1
\, . \nonumber
\end{align}
The latter will be relevant for establishing the form of the cyclic permutation form of the $\chi_5$ contribution to the one-loop heptagon.


\end{document}